\documentclass[12pt]{iopart}
\usepackage{cite}
\begin{document}
\title[Quantum correlations ] {Quantum correlations expressed as information
and entropic inequalities for composite and noncomposite systems }
\author{Margarita A Man'ko and Vladimir I Man'ko}
\address{P N Lebedev Physical Institute, Leninskii Prospect 53,
Moscow 119991, Russia} \ead{mmanko@sci.lebedev.ru} \ead{manko@sci.lebedev.ru}
\begin{abstract}
For noncomposite systems in classical and quantum domains, we obtain new
inequalities such as the subadditivity and strong subadditivity conditions for
Shannon entropies and information determined by the probability distributions
and for von Neumann entropies of quantum states determined by the density
operators. We extend the relations of Shannon and Tsallis entropies to the
entropies of conditional probability distributions known for composite systems
to the case of noncomposite systems. We give a review of the approach to
construct the tomographic-probability distributions for qudit systems and
present the entropic and information inequalities for spin tomograms, as well
as the subadditivity and strong subadditivity conditions for tomograms of the
both noncomposite and composite system states.
\end{abstract}







 \pacs{03.65.-w, 03.65.Ta, 02.50.Cw, 03.67.-a.}


\section{Introduction}
Quantum correlations are the properties of quantum systems which can be
associated with different phenomena. The correlations are determined by the
structure of the density operator~\cite{Landau,vonNeumann}. The density
operator of both composite and noncomposite quantum systems has only
nonnegative eigenvalues, and this property provides the constraints
determining the domain of parameters associated with matrix elements of the
state density matrices in arbitrary representations. For composite systems,
the density operator has different structure with respect to the decomposition
in the form of convex sum of direct products of the density operators of the
subsystem states.

The entanglement~\cite{Schrod26} of composite quantum systems containing two
or more subsystems is one of the important manifestations of quantum
correlations. The entanglement for two qubits can be detected by violation of
the Bell inequality~\cite{Bell,HornClauser}.

Recently~\cite{Mancini96}, it was demonstrated that the density operators can
be mapped onto the probability vectors (quantum tomographic probabilities
called state tomograms), which determine the operators. Recent reviews of this
approach can be found in \cite{NuovoCim,VovaJETP}. Information and entropic
properties of quantum systems were studied in the tomographic-probability
representation in quantum mechanics in \cite{RuiJRLR,Beauty,RitaPS13,ELZE}.

The other property reflecting the presence of quantum correlations in
composite systems is quantum discord~\cite{Zurek,Vedral,YurkevichJRLR,Isar};
it corresponds to a difference in information properties of bipartite
classical and quantum systems. The discord shows a difference between the
properties of the Shannon entropy~\cite{Shannon} (determined by the
probability distribution describing the classical state of a composite system)
and the von Neumann entropy of the quantum bipartite system. Quantum
correlations in noncomposite quantum systems were found to be described by the
contextuality phenomenon~\cite{Kochen,Shumo,Vourdas,Cabello,Strakhov}.

Any probability distribution, in addition to the Shannon entropy, determines
the so-called $q$-entropies~\cite{Renyi,Tsallis} containing an extra parameter
dependence. The $q$-entropies are also defined for quantum states, and they
are determined by the state density operator.

The Shannon and $q$-entropies obey some inequalities for composite systems in
both classical and quantum domains. The strong subadditivity condition for the
von Neumann entropy of three-partite system was proved in \cite{LiebRuskai}.
There are other aspects of the entropic inequalities studied in
\cite{Ruskai,Lieb,6,4,5,8,7}.

An entropic inequality called the subadditivity condition for the Shannon
entropy of the classical bipartite system and the von Neumann entropy of the
quantum bipartite system corresponds to the degree of correlations of the
subsystem degrees of freedom.
Recently~\cite{JRLRVova,JRLR-3-13-M+V,PST-13-M+V}, it was shown that analogous
subadditivity conditions and information also exist for noncomposite systems,
both classical and quantum. The subadditivity condition and information
correspond to the degree of intrinsic correlations between different groups of
the results obtained in the experiments where the classical and quantum
observables are measured. An analogous result about the existence of the
strong subadditivity condition for noncomposite classical and quantum systems
was obtained in \cite{PST-13-M+V}.

The aim of this paper is to review the new entropic and information
inequalities found for noncomposite systems
following~\cite{JRLRVova,JRLR-3-13-M+V,PST-13-M+V} and introduce for these
systems new concepts -- analogs of the conditional entropies associated with
composite systems and obtain new equalities characterizing the conditional
entropies for Shannon entropy and $q$-entropies.

This paper is organized as follows.

In section~2, we review the properties of sets of nonnegative numbers
organized as tables with different labels by collective indices and obtain
various inequalities for these nonnegative numbers. In section~3, we study new
entropic inequalities for tomographic entropies of qudit states. In section~4,
we present new entropic inequalities for density matrices of noncomposite
systems, while in section~5 we discuss the quantum discord and its properties
for noncomposite systems. We give our conclusions and prospectives in
section~6.

\section{Tables of nonnegative numbers}
In this section, we describe a tool how to consider any set of $N$ nonnegative
numbers, the sum of which is equal to unity, as a set analogous to a vector or
a set considered as a matrix. Also we show how such a set can be viewed as a
table where each position in the table is labeled by three integers, or four
integers, etc. The inverse procedure can also be suggested. This means that an
arbitrary table which contains nonnegative numbers can be viewed as a vector
with nonnegative number components. This vector can be identifies with the
probability vector or a point in the simplex. Also the table, where the
nonnegative numbers are situated in the positions labeled by some collection
of integers, can be treated as a joint probability distribution. When we
identify the table with nonnegative numbers with the joint probability
distribution, we take into account an additional information arising from the
results of experiments where some random variables are measured.

As it was pointed out in \cite{JRLR-3-13-M+V} and employed in
\cite{PST-13-M+V,JRLRVova}, there exist some entropic inequalities which can
be associated not only with the probability distributions but with the sets of
nonnegative numbers independently on the fact that these numbers are used as
probabilities. Also in \cite{PST-13-M+V}, it was stated that some properties
of the von Neumann entropy can be considered as the properties of the complex
number set where the complex numbers are associated with matrix elements of
the quantum-state density matrix. Thus, in this paper we concentrate on the
question: What inequalities are completely mathematical characteristics of
sets of either nonnegative numbers or complex numbers, and what inequalities
incorporate information that the numbers are associated with physical
characteristics of the classical or quantum systems?

Thus, we can consider $N$ nonnegative numbers and label these numbers by the
integer number $k=1,2,\ldots,N$. We obtain a set of nonnegative numbers $\vec
p=(p_1,p_2,\ldots,p_N)$, where $p_k\geq 0$ and $\sum_{k=1}^Np_k=1$, and the
vector notation $\vec p$ is introduced. If we add $M$ zero numbers to this
set, we get the set of $N'=N+M$ nonnegative numbers; the number $M$ can be
chosen arbitrary. This means that one can construct another vector $\vec p'$
with $N'$ components. Some of these components are equal to zero.

Let us consider the same initial set of $N$ nonnegative numbers. We can
introduce a different labeling of these numbers if, e.g., the chosen integer
has the form $N=N_1N_2$, where $N_1$ and $N_2$ are also integers. Then we
organize these numbers as a table with labels $(jk)$. Thus, we obtain  ${\cal
P}_{kj}$, where $k=1,2,\ldots, N_1$ and $j=1,2,\ldots, N_2$, satisfying
$\sum_{k=1}^{N_1}\sum_{j=1}^{N_2}{\cal P}_{kj}=1$. The numbers $N_1$ and $N_2$
can be chosen in a different way, if the number $N=n_1n_2\cdots n_s$ with
factors $n_\alpha$ $(\alpha=1,2,\ldots,s)$ equal to integers. In fact, we can
combine the integers in this product, e.g., as $n_1=N_1$, $n_2n_3\cdots
n_s=N_2$. Thus, we can map the initial set of $N$ nonnegative numbers on
anyone of the tables ${\cal P}_{jk}$ constructed according to the described
representation of the integer $N$ in the form of the product of the integers.
If the number $N$ is prime number, one can add zeros and consider the integer
$N'$ for which one has the representation in the form of product of integers.
Thus, one can always find many tables ${\cal P}_{jk}$, where indices $(jk)$
play the role of positions, and the index $\alpha$ labels a particular table
associated, e.g., with different numbers $N'$.

The consideration above presented demonstrates that any set of nonnegative
numbers can be mapped onto many tables (matrices ${\cal P}_{jk}$). These
rectangular matrices have matrix elements equal either to zero or to a number
$p_k$ from the initial vector $\vec p$. One can continue this construction of
maps to associate the initial $N$ nonnegative numbers $p_k$ with, e.g., the
table $\Pi_{kjl}$ $(k=1,2,\ldots N_1$, $j=1,2,\ldots N_2$, and $l=1,2,\ldots
N_3)$. The positions in this table are labeled by three integers. We used the
decomposition $N=N_1N_2N_3$ of integer $N$ into product of three integers. As
it is clear from the previous discussion, such a decomposition can always be
found by constructing the integer $N'=N+M$ and using the table containing $N'$
numbers. $N$ numbers in this table $\Pi_{kjl}$ are the initial $N$ numbers
$p_k$, and the other numbers are zeros.

It is clear that there exist many possibilities to construct different tables
$\Pi_{kjl}^{(\alpha)}$ depending on the choice of the integer $M$ and integers
$N_1$, $N_2$, and $N_3$ providing the decomposition $N'=N_1N_2N_3$. The index
$\alpha $ labels different tables. It is also clear that for any initial set
of nonnegative number $p_k$ one can construct the map of these numbers onto
table $T_{j_1,j_2,\ldots,j_k}$ containing all these numbers and the
corresponding number of zero components. If the nonnegative numbers are
experimentally measured probabilities, the tables constructed can be
associated with probability distributions. The chosen table ${\cal P}_{kj}$ of
nonnegative numbers provides two sets ${\cal P}_{1k}$ and ${\cal P}_{2j}$ of
nonnegative numbers
\begin{equation}\label{3}
P_{1k}=\sum_{j=1}^{N_2}{\cal P}_{kj},\quad P_{2j}=\sum_{k=1}^{N_1}{\cal
P}_{kj}.
\end{equation}
Also the chosen table of nonnegative numbers $\Pi_{kjl}$ provides the three
other tables of nonnegative parameters
\begin{equation}\label{7}
 {\cal P}_{kj}^{(12)}=\sum_{l=1}^{N_3}\Pi_{kjl},\quad{\cal
P}_{jl}^{(23)}=\sum_{k=1}^{N_1}\Pi_{kjl},\quad 
P_{j}^{(2)}=\sum_{k=1}^{N_1}\sum_{l=1}^{N_3}\Pi_{kjl}.
\end{equation}
If the considered tables of nonnegative numbers are identified with joint
probability distributions of physical composite systems, the above formulas
correspond to constructing marginal distributions for bipartite and tripartite
systems, respectively.

The constructed map of a set of nonnegative numbers $p_k$ onto different kinds
of tables, like ${\cal P}_{kj}$ or $\Pi_{kjl}$, etc., has an inverse in the
following sense.

For a given table with $N_1N_2$ positions (matrix ${\cal P}_{kj})$, one can
construct $N$-vector using the map of pairs of integers onto integers as
$(11)\to 1,(12)\to 2,\ldots,(1N_2)\to N_2,(2,1)\to N_2+1,(2,2)\to
N_2+2,\ldots,(N_1N_2)\to N$. Analogously, the table of nonnegative numbers
$\Pi_{kjl}$ can be mapped onto the $N$-vector. This observation can be used to
rewrite the relations available for tables ${\cal P}_{kj}$ or $\Pi_{kjl}$ in
terms of numbers $p_k$. For example, it is known that joint probability
distributions provide inequalities called the subadditivity conditions and
strong subadditivity conditions, and these inequalities are valid for matrices
${\cal P}_{kj}$ and joint probability distributions $\Pi_{kjl}$, respectively.
Being rewritten in the form of probability vector components $p_k$, they are
valid for an arbitrary set of nonnegative numbers, which are equal to the
probabilities.

\newpage

\subsection*{Example of ${N=7}$}
For $N=7$, one has the inequalities~\cite{PST-13-M+V}
\begin{eqnarray}
\fl\left(-\sum_{k=1}^{7}p_{k}\ln
p_{k}\right)-(p_1+p_2+p_5+p_6)\ln(p_1+p_2+p_5+p_6)\nonumber\\
\fl-(p_3+p_4+p_7)\ln(p_3+p_4+p_7)
\leq -(p_1+p_2)\ln(p_1+p_2)-(p_3+p_4)\ln(p_3+p_4)\nonumber\\
\fl-(p_5+p_6)\ln(p_5+p_6)-p_7\ln
p_7-(p_1+p_5)\ln(p_1+p_5)-(p_2+p_6)\ln(p_2+p_6)\nonumber\\
\fl -(p_3+p_7)\ln(p_3+p_7)-p_4\ln p_4\label{12}
\end{eqnarray}
and
\begin{eqnarray}
&&-\sum_{k=1}^{7}p_{k}\ln p_{k}\leq
-(p_1+p_2+p_5+p_6)\ln(p_1+p_2+p_5+p_6)\nonumber\\
&&-(p_3+p_4+p_7)\ln(p_3+p_4+p_7)-(p_1+p_3)\ln(p_1+p_3)\nonumber\\
&&-(p_2+p_4)\ln(p_2+p_4)-(p_5+p_7)\ln(p_5+p_7)-p_4\ln p_4.\label{13}
\end{eqnarray}
These inequalities are analogs of the strong subadditivity condition for
three-partite system and the subadditivity condition for two-partite system,
respectively. In spite of the fact that the 7 nonnegative numbers are not
associated with any probabilities, the inequalities are satisfied. The
inequalities were obtained by introducing an $N'$-dimensional vector with
$N'=N+1$ with the component $p_8=0$. The eight-dimensional vector obtained was
mapped onto the rectangular matrix ${\cal P}_{kj}$ corresponding to the
decomposition $N'=2\cdot 4$, and this provided the subadditivity
condition~({\ref{13}). The strong subadditivity condition~({\ref{12})
corresponds to the decomposition $N'=2\cdot 2\cdot 2$.

For four nonnegative numbers $\vec p=(a,b,c,d)$, an analog of the
subadditivity condition discussed in \cite{{JRLR-3-13-M+V}} reads
\begin{eqnarray}\label{M3}
\fl-(a+b)\ln(a+b)-(c+d)\ln(c+d)-(a+c)\ln(a+c)-(b+d)\ln(b+d)\nonumber\\
\geq-a\ln a-b\ln b-c\ln c-d\ln d.
\end{eqnarray}

\subsection*{Inequalities for tables}
The inequalities above discussed are examples of general inequalities for
matrix tables ${\cal P}_{kj}$ and $\Pi_{kjl}$ containing the nonnegative
numbers; they are
\begin{eqnarray}
\fl-\sum_{j=1}^m\left(\sum_{k=1}^n{\cal
P}_{jk}\right)\left(\ln\sum_{k'=1}^n{\cal P}_{jk'}\right)
-\sum_{k=1}^n\left(\sum_{j=1}^m{\cal P}_{jk}\right)\left(\ln\sum_{j'=1}^n{\cal P}_{jj'}\right)\nonumber\\
\geq
-\sum_{j=1}^m\sum_{k=1}^n{\cal P}_{jk}\ln {\cal P}_{jk},\label{M1}\\
\fl-\sum_{j=1}^{n_1}\sum_{k=1}^{n_2}\sum_{m=1}^{n_3}\Pi_{jkm}\ln \Pi_{jkm}-
\sum_{k=1}^{n_2}\left(\sum_{j=1}^{n_1}\sum_{m=1}^{n_3}\Pi_{jkm}\right)
\ln \left(\sum_{j'=1}^{n_1}\sum_{m'=1}^{n_2}\Pi_{j'km}\right)\nonumber\\
\fl\leq-\sum_{j=1}^{n_1}\sum_{k=1}^{n_2}\left(\sum_{m'=1}^{n_3}\Pi_{jkm'}\right)
\ln \left(\sum_{m=1}^{n_3}\Pi_{jkm}\right)
-\sum_{k=1}^{n_2}\sum_{m=1}^{n_3}\left(\sum_{j'=1}^{n_1}\Pi_{j'km}\right) \ln
\left(\sum_{j=1}^{n_1}\Pi_{jkm}\right). \label{M18}
\end{eqnarray}
In the case where ${\cal P}_{kj}$ is the joint probability distribution for
two random variables of the bipartite system, equation~(\ref{M1}) is the usual
subadditivity condition for Shannon entropies associated with the probability
distribution. Inequality~(\ref{M18}) is the strong subadditivity condition, if
$\Pi_{jkm}$ is the joint probability distribution of three random variables
for the tripartite system. One should stress that the inequalities discussed
are valid for an arbitrary probability vector $\vec p$ even of the system
which is not composite. Such a system can have only a single random variable;
nevertheless, there exist the inequalities for the probability distribution of
this random variable identical to the subadditivity and strong subadditivity
conditions. For the systems, which do not have subsystems, these inequalities
can be interpreted as entropic properties corresponding to the existence of
correlations between different results of measurements of the random variable.

\section{Inequalities for tomographic entropies}
For qudit states with the density matrix $\rho$, the tomographic-probability
distribution (spin tomogram) $w(m,\vec n)$ can be used as an alternative to
the density matrix $\rho$~\cite{DodPLA,OlgaJETP}. It is interpreted as the
probability to obtain the spin projection $m$, where $m=-j,-j+1,\ldots,j$, on
the quantization axes determined by the unit vector $\vec
n=(\sin\theta\cos\varphi,\sin\theta\sin\varphi,\cos\theta)$. The tomogram can
be obtained as a result of the experiment. Thus, for qudit states, nonnegative
numbers $p_k$ can be identified with the tomographic probabilities. For
example, for the qudit state corresponding to spin $j=3/2$ we have the
four-vector $\vec p=(p_1,p_2,p_3,p_4)$, where $p_1=w(-3/2,\vec n)$,
$p_2=w(-1/2,\vec n)$, $p_3=w(1/2,\vec n)$, and $p_4=w(3/2,\vec n)$.

The tomographic Shannon entropy is defined as (see, e.g., \cite{RitaFP})
\begin{equation}\label{Shannon}
H(\vec n)=-\sum_{m=-3/2}^{3/2}w(m,\vec n)\ln w(m,\vec n).
\end{equation}
There are two tomographic-probability distributions
\begin{eqnarray}
\fl W_1(-2,\vec n)=w(-3/2,\vec n)+w(-1/2,\vec n), \qquad W_1(2,\vec
n)=w(1/2,\vec
n)+w(3/2,\vec n),\label{W1}\\
\fl W_2(-1,\vec n)=w(-3/2,\vec n)+w(1/2,\vec n), \qquad W_2(1,\vec
n)=w(-1/2,\vec n)+w(3/2,\vec n),\label{W1}\end{eqnarray} which are analogs of
marginal probability distributions for the bipartite system. We can interpret
these probability distributions as follows: $W_1(\vec n)$ yields the
probabilities to find the qudit state with the sum of spin projections equal
to $\pm 2$, and $W_2(\vec n)$ provides the probabilities to find the qudit
state with the sum of spin projections equal to $\pm 1$. In spite the fact
that the system is not bipartite, we have the subadditivity condition
\begin{eqnarray}
\fl -\sum_{m=-3/2}^{3/2}w(m,\vec n)\ln w(m,\vec n)\leq - W_1(1,\vec n)\ln
W_1(1,\vec n)-W_1(-1,\vec n)\ln W_1(-1,\vec n)\nonumber\\
-W_2(2,\vec n)\ln
W_2(2,\vec n)-W_2(-2,\vec n)\ln W_2(-2,\vec n).\label{subadd} \end{eqnarray}

\subsection*{Conditional tomographic probabilities}
 We introduce the conditional tomographic probability distribution
for the qudit state following the procedure of constructing the distribution
for the bipartite system. For the qudit-state tomogram with $j=3/2$, we
introduce two conditional probabilities
\begin{eqnarray}
\fl V_1(\vec n)=\frac{w(-3/2,\vec n)}{w(-3/2,\vec n)+w(-1/2,\vec n)}\,,\qquad
V_2(\vec n)=\frac{w(-1/2,\vec n)}{w(-3/2,\vec n)+w(-1/2,\vec
n)}\,,\label{VV}\\
\fl \widetilde V_1(\vec n)=\frac{w(1/2,\vec n)}{w(3/2,\vec n)+w(1/2,\vec
n)}\,,\qquad \widetilde V_2(\vec n)=\frac{w(3/2,\vec n)}{w(3/2,\vec
n)+w(1/2,\vec n)}\,.\label{tildeVV} \end{eqnarray} The tomographic conditional
probability distributions~(\ref{VV}) and (\ref{tildeVV}) determine the
tomographic conditional entropies. Since there exists the identity for four
nonnegative numbers $p_1$, $p_2$, $p_3$, and $p_4$ (considered as the
probabilities) of the form
\begin{eqnarray}
-p_1\ln\frac{p_1}{p_1+p_2}-p_2\ln\frac{p_2}{p_1+p_2}-p_3\ln\frac{p_3}{p_3+p_4}-p_4\ln\frac{p_4}{p_3+p_4}\nonumber\\
\fl =(p_1+p_2)\ln(p_1+p_2)+(p_3+p_4)\ln(p_3+p_4)-p_1\ln p_1-p_2\ln p_2-p_3\ln
p_3-p_4\ln p_4,\nonumber\\
\label{pppp}
\end{eqnarray}
one can rewrite this equality as the property of the conditional tomographic
entropy
\begin{equation}\label{ConEnt}
H(V,\widetilde V)=H(V\mid\widetilde V)+H(\widetilde V),
\end{equation}
where the tomographic entropy of the qudit state reads
\begin{equation}\label{TomEnt}
H(V,\widetilde V)=-\sum_{m=-3/2}^{3/2}w(m,\vec n)\ln w(m,\vec n).
\end{equation}
Formula~(\ref{ConEnt}) can be generalized to yield an analog of the chain form
of the conditional entropic equality for multipartite system.

The conditional tomographic entropy of the qudit state is defined as
\begin{equation}\label{ConTomEnt}
\fl H(V\mid\widetilde V)=H(V,\widetilde V)+W_1(2,\vec n)\ln W_1(2,\vec
n)+W_1(-2,\vec n)\ln W_1(-2,\vec n).
\end{equation}

\subsection*{Conditional $q$-entropies}

We define the $q$-entropy for four nonnegative numbers $p_1$, $p_2$, $p_3$,
and $p_4$ as follows:
\begin{equation}\label{T1}
T_q(\vec p)=\frac{1}{1-q}\,\Big( \sum_{k=1}^4p_k^q-1\Big). \end{equation}
Identifying the numbers $p_k$ with tomographic probabilities of qudit state
with $j=3/2$, we obtain the Tsallis tomographic entropy~\cite{RitaFP}
\begin{equation}\label{Ts}
T_q(\vec n)=\frac{1}{1-q}\,\Big( \sum_{m=-3/2}^{3/2}[w(m,\vec n)]^q-1\Big).
\end{equation}
For $q\to 1$, this entropy tends to the conditional Shannon entropy
$H(V\mid\widetilde V)$.

One can introduce the conditional tomographic $q$-entropy $T_q(V\mid
\widetilde V)$ for the qudit state with $j=3/2$, using the formula presented
in \cite{JRLR-3-13-M+V} for bipartite qudit states,
\begin{equation}\label{T3}
T_q(V\mid \widetilde V)=T_q(\vec n)-\frac{1}{1-q}\,\Big(W_1(-2,\vec n)^q+
W_1(2,\vec n)^q-1\Big).
\end{equation}
Formula~(\ref{T3}) can be generalized to yield the chain form of $q$-entropic
equality for the conditional entropies associated with multipartite systems.

In view of the inequalities derived for tomographic $q$-entropies of bipartite
qudit systems in \cite{JRLR-3-13-M+V}, we obtain analogous inequalities for
$q$-entropy of the qudit state under consideration
\begin{equation}\label{T4}
T_q(V)\leq T_q(\vec n),\qquad T_q(V\mid \widetilde V)\leq T_q(V),
\end{equation}
where $$ 
T_q(V)=\frac{1}{1-q}\Big(W_1(-2,\vec n)^q+ W_1(2,\vec n)^q-1\Big),\quad
T_q(\widetilde V)=\frac{1}{1-q}\Big(W_2(1,\vec n)^q+ W_2(-1,\vec
n)^q-1\Big). $$ 

\section{Inequalities for density matrices}
The quantum states of qudits with spin $j$ $(2j+1=N)$ are determined by
density $N$$\times$$N$-matrices $\rho_{km}$. We can apply our tool elaborated
for the set of complex numbers $\rho_{km}$. In this case, the tool is to map a
pair of integers $(k,m)$ onto the other collective pairs of integers, e.g.,
$k\to(l_k,j_k)$ and $m\to(l_m,j_m)$, i.e., $(k,m)\to(l_kj_k,l_mj_m)$.
Continuing this procedure, we construct the map
$(k,m)\to(l_kj_kn_k,l_mj_mn_m)$, etc. This means that the density matrix
$\rho_{mm'}$ with $j=3/2$ can be written in the form
\begin{equation}\label{DM}
\rho=\left( \begin{array}{cccc}
\rho_{11,11}&\rho_{11,12}&\rho_{11,21}&\rho_{11,22}\\
\rho_{12,11}&\rho_{12,12}&\rho_{12,21}&\rho_{12,22}\\
\rho_{21,11}&\rho_{21,12}&\rho_{21,21}&\rho_{21,22}\\
\rho_{22,11}&\rho_{22,12}&\rho_{22,21}&\rho_{22,22}
\end{array} \right).
\end{equation}
Such a form provides a possibility to apply to this density matrix the
positive maps which yield the nonnegative matrices
\begin{equation}\label{DM1}
\rho(1)_{jk}=\sum_{s=1}^2\rho_{js,ks},\qquad\rho(2)_{mn}=\sum_{s=1}^2\rho_{sm,sn}.
\end{equation}
Complex numbers $\rho_{mm'}$ satisfy the criterion of nonnegativity of the
matrix $\rho$, as well as the complex numbers $\rho(1)_{jk}$ and
$\rho(2)_{mn}$. Then we have the subadditivity condition
\begin{equation}\label{DM2}
-\mbox{Tr}\,\rho\ln\rho\leq -\mbox{Tr}\,\rho(1)\ln\rho(1)-
\mbox{Tr}\,\rho(2)\ln\rho(2).
\end{equation}
This inequality is valid for one qudit state, and the qudit is a system which
does not contain subsystems.

\subsection*{Examples of spin-2 and spin-3 states}

There are other examples~\cite{JRLR-3-13-M+V} of such inequalities which are
analogs of the strong subadditivity condition for qudit states with $j=2$ and
$j=3$. In the case of 7$\times$7-matrix $\rho_{jk}$, it has the form
\begin{equation}\label{SSC1}
-\mbox{Tr}\left(\rho\ln\rho\right)-\mbox{Tr}\left(R_2\ln R_2\right)\leq
-\mbox{Tr}\left(R_{12}\ln R_{12}\right)-\mbox{Tr}\left(R_{23}\ln
R_{23}\right),
\end{equation}
where the density matrix $R_{12}$ has matrix elements expressed in terms of
the density matrix $\rho_{jk}$ as follows:
\begin{equation}\label{SSC12} R_{12}=
\pmatrix{\rho_{11}+\rho_{22}&\rho_{13}+\rho_{24}&\rho_{15}+\rho_{26}&\rho_{17}\cr
\rho_{31}+\rho_{42}&\rho_{33}+\rho_{44}&\rho_{35}+\rho_{46}&\rho_{37}\cr
\rho_{51}+\rho_{62}&\rho_{53}+\rho_{64}&\rho_{55}+\rho_{66}&\rho_{57}\cr
\rho_{71}&\rho_{73}&\rho_{75}&\rho_{77}\cr}.\end{equation} The density matrix
$R_{23}$ reads
\begin{equation}\label{SSC23} R_{23}=
\pmatrix{\rho_{11}+\rho_{55}&\rho_{12}+\rho_{56}&\rho_{13}+\rho_{57}&\rho_{14}\cr
\rho_{21}+\rho_{65}&\rho_{22}+\rho_{66}&\rho_{23}+\rho_{67}&\rho_{24}\cr
\rho_{31}+\rho_{75}&\rho_{32}+\rho_{76}&\rho_{33}+\rho_{77}&\rho_{34}\cr
\rho_{41}&\rho_{42}&\rho_{43}&\rho_{44}\cr},\end{equation} while the matrix
$R_{2}$ is
\begin{equation}\label{SSC2} R_2=
\pmatrix{\rho_{11}+\rho_{22}+\rho_{55}+\rho_{66}&\rho_{13}+\rho_{24}+\rho_{57}\cr
\rho_{31}+\rho_{42}+\rho_{75}&\rho_{33}+\rho_{44}+\rho_{77}\cr}.\end{equation}

In the case of 5$\times$5-matrix $\rho_{lk}$ corresponding to the qudit state
with $j=2$, we have inequality~(\ref{SSC1}), where
\begin{eqnarray}\fl R_{12}=
\pmatrix{\rho_{11}+\rho_{22}&\rho_{13}+\rho_{24}&\rho_{15}\cr
\rho_{31}+\rho_{42}&\rho_{33}+\rho_{44}&\rho_{35}\cr
\rho_{51}&\rho_{53}&\rho_{55}\cr},\qquad R_{23}=
\pmatrix{\rho_{11}+\rho_{55}&\rho_{12}&\rho_{13}&\rho_{14}\cr
\rho_{21}&\rho_{22}&\rho_{23}&\rho_{24}\cr
\rho_{31}&\rho_{32}&\rho_{33}&\rho_{34}\cr
\rho_{41}&\rho_{42}&\rho_{43}&\rho_{44}\cr},\label{SSC12-2}\\  R_2=
\pmatrix{\rho_{11}+\rho_{22}+\rho_{55}&\rho_{13}+\rho_{24}\cr
\rho_{31}+\rho_{42}&\rho_{33}+\rho_{44}\cr}.\label{SSC2-2}
\end{eqnarray}
In order to obtain these inequalities, we mapped the matrices $\rho_{lk}$ on
matrices $R_{lk}$ acting on vectors in the eight-dimensional linear space. The
matrix $R_{lk}$ has matrix elements of the matrix $\rho$, and the other matrix
elements are zeros.

The quantum subadditivity condition for qutrit state was obtained in
\cite{JRLRVova}. The procedure developed is analogous to mapping of
$N$-vectors on $N'$-vectors with adding zero components to the initial vector.
After this map, we apply the known subadditivity and strong subadditivity
conditions to the obtained nonnegative matrix $R_{lk}$. As a result, we get
the above entropic inequalities for the density matrices of qudit states. It
is obvious that analogous generalized inequalities can be obtained for
arbitrary qudits with $j=3/2,2,5/2,3,7/2,\ldots$ Also the strong subadditivity
condition, which is usually associated with three-partite systems, can be
written for the two-qudit system, in view of the suggested map of integers on
the collection of integers.

\section{Some inequalities and quantum discord}
For a bipartite system with the joint probability distribution ${\cal
P}_{jk}$, the Shannon information is defined as a difference of the sum of
subsystem entropies and entropy of the composite system. The two-qudit state
can be described by the joint tomographic probability distribution
$w(m_1,m_2,u)$, which is given by diagonal matrix elements of the density
matrix $\rho(1,2)$ in unitary rotated basis
\begin{equation}\label{20}
w(m_1,m_2,u)=\langle m_1m_2\mid u^\dagger\rho(1,2)u\mid m_1m_2\rangle,
\end{equation}
where $u$ is the unitary matrix.

In the case $u=u_1\otimes u_2$, where $u_1$ and $u_2$ are matrices of the
irreducible representation of the rotation group, the tomogram $w(m_1,m_2,\vec
n_1,\vec n_2)$ is the joint probability distribution of two random spin
projections $m_1$ and $m_2$. The directions $\vec n_1$ and $\vec n_2$ are
determined by Euler angles which fix the matrix elements of the representation
matrices $u_1$ and $u_2$. The Shannon entropy
\begin{equation}\label{22}
H_{12}(u)=-\sum_{m_1=-j_1}^{j_1}\sum_{m_2=-j_2}^{j_2} w(m_1,m_2,u)\ln
w(m_1,m_2,u)
\end{equation}
has the minimum for a specific value of the unitary matrix $u$~\cite{RuiJRLR},
which is equal to the von Neumann entropy associated with the density matrix
of the two qudit states, i.e.,
\begin{equation}\label{22min}
\min H(u)=-\mbox{Tr}\,\rho(1,2)\ln \rho(1,2).
\end{equation}
The density matrices $\rho(1)=-\mbox{Tr}_2\,\rho(1,2)$ and
$\rho(2)=-\mbox{Tr}_1\,\rho(1,2)$ of the qudits give the von  Neumann
entropies $S(1)=-\mbox{Tr}\,\rho(1)\ln\rho(1)$ and
$S(2)=-\mbox{Tr}\,\rho(2)\ln\rho(2)$. These entropies can be found as minima
of the Shannon entropies
\begin{equation}\label{26}
H_k(u)=-\sum_{m_k=-j_k}^{j_k}w_k(m_k,u)\,\ln w_k(m_k,u), \qquad k=1,2
\end{equation}
with respect to local unitary transforms $u=u_{10}\otimes u_{20}$. The
subadditivity condition reads
\begin{equation}\label{27}
S_{12}\leq S_{1}+S_{2}.
\end{equation}
In \cite{JRLR-3-13-M+V}, we showed that inequality~(\ref{27}) can be written
as
\begin{equation}\label{E}
S_{1}+S_{2}\geq H_{12}(u_{10}\otimes u_{20})\geq S_{12}.
\end{equation}
The tomographic discord introduced as the difference
\begin{equation}\label{G}
{\cal D}=(S_{1}+S_{2}-S_{12})-I(u_{10}\otimes u_{20})\geq 0
\end{equation}
characterizes quantum correlations in the bipartite system of qudits.

\subsection*{Quantum discord for spin-3/2 state}

In view of the inequalities discussed above, we can introduce the quantum
discord for one qudit state. For example, the qudit state corresponding to
$j=3/2$ with density matrix~(\ref{DM}) determines the state tomogram
$w(m,u)=\langle m\mid u\rho u^\dagger\mid m\rangle$, $m=-3/2,-1/2,1/2,3/2$.
The von Neumann entropy of this state is the minimum of the Shannon entropy
determined by the tomographic probability distribution; this means that
$\min_u\Big(-\sum_{m=-3/2}^{3/2} w(m,u)\Big)=-\mbox{Tr}\,\rho\ln\rho$. We
rewrite the matrix in two forms: one form corresponds to an arbitrary matrix,
and the other one specifies the spin-3/2 state,
\begin{equation}\label{newmat1}
\fl\rho= \pmatrix{\rho_{11}&\rho_{12}&\rho_{13}&\rho_{14}\cr
\rho_{21}&\rho_{22}&\rho_{23}&\rho_{24}\cr
\rho_{31}&\rho_{32}&\rho_{33}&\rho_{34}\cr
\rho_{41}&\rho_{42}&\rho_{43}&\rho_{44}\cr}\equiv
\pmatrix{\rho_{-3/2~-3/2}&\rho_{-3/2~-1/2}&\rho_{-3/2~1/2}&\rho_{-3/2~3/2}\cr
\rho_{-1/2~-3/2}&\rho_{-1/2~-1/2}&\rho_{-1/2~1/2}&\rho_{-1/2~3/2}\cr
\rho_{1/2~-3/2}&\rho_{1/2~-1/2}&\rho_{1/2~1/2}&\rho_{-1/2~3/2}\cr
\rho_{3/2~-3/2}&\rho_{3/2~-1/2}&\rho_{3/2~1/2}&\rho_{3/2~3/2}\cr}.\end{equation}
We consider two density matrices $\rho(1)$ and $\rho(2)$ by taking analogs of
tracing with respect to subsystem degrees of freedom; the 2$\times$2-matrices
read
\begin{eqnarray} \rho(1)=
\pmatrix{\rho_{11}+\rho_{22}&\rho_{13}+\rho_{24}\cr
\rho_{31}+\rho_{42}&\rho_{33}+\rho_{44}\cr},\qquad \rho_{2}=
\pmatrix{\rho_{11}+\rho_{33}&\rho_{12}+\rho_{34}\cr
\rho_{21}+\rho_{43}&\rho_{22}+\rho_{44}\cr}.\label{newmat2}
\end{eqnarray}
We constructed the 2$\times$2-matrices from arbitrary density
4$\times$4-matrices. In notation relevant to the spin-3/2 state, these density
matrices read
\begin{eqnarray}\label{newmat3}
\rho^{(3/2)}(1)=
\pmatrix{\rho_{-3/2~-3/2}+\rho_{-1/2~-1/2}&\rho_{-3/2~1/2}+\rho_{-1/2~3/2}\cr
\rho_{-1/2~-3/2}+\rho_{-3/2~-1/2}&\rho_{1/2~1/2}+\rho_{3/2~3/2}\cr},\nonumber\\[-2mm]
\\[-2mm]
\rho^{(3/2)}(2)=
\pmatrix{\rho_{-3/2~-3/2}+\rho_{1/2~-1/2}&\rho_{-3/2~-1/2}+\rho_{-1/2~3/2}\cr
\rho_{-1/2~-3/2}+\rho_{3/2~-1/2}&\rho_{-1/2~-1/2}+\rho_{3/2~3/2}\cr};\nonumber
\end{eqnarray}
they determine the tomograms. While deriving (\ref{newmat3}), we used the
other notation for the matrices $\rho^{(3/2)}(1)= \pmatrix{r_{11}&r_{12}\cr
r_{21}&r_{22}\cr}$ and $\rho^{(3/2)}(2)= \pmatrix{R_{11}&R_{12}\cr
R_{21}&R_{22}\cr}$. The two tomograms $w(\alpha, u_1)$ and $w(\beta, u_2)$ are
diagonal matrix elements of the matrices $u_1\rho^{(3/2)}(1)u_1^\dagger$ and
$u_2\rho^{(3/2)}(2)u_2^\dagger$, they are
\begin{equation}\label{newmat4}
\fl w(\alpha, u_1)=\langle \alpha\mid
u_1\rho^{(3/2)}(1)u_!^\dagger\mid\alpha\rangle,\qquad w(\beta, u_2)=\langle
\beta\mid u_2\rho^{(3/2)}(2)u_2^\dagger\mid\beta\rangle.
\end{equation}
The random variables $\alpha$ and $\beta$ take two values $\pm 1$. The von
Neumann entropies of the ``qubit'' states with density matrices
$\rho^{(3/2)}(1)$ and $\rho^{(3/2)}(2)$ read
\begin{equation}\label{newmat5}
\fl S^{(3/2)}(1)=-\mbox{Tr}\,\rho^{(3/2)}(1)\ln\rho^{(3/2)}(1),\qquad
S^{(3/2)}(2)=-\mbox{Tr}\,\rho^{(3/2)}(2)\ln\rho^{(3/2)}(2). \end{equation} We
introduce quantum discord as follows:
\begin{equation}\label{newmat6}
 D=\mbox{Tr}\,(\rho\ln\rho)+S^{(3/2)}(1)+S^{(3/2)}(2)-I(u_{10}\times u_{20}),
\end{equation}
where the tomographic information $I(u)$ is
\begin{eqnarray}\label{newmat7}
I(u)=\sum_{m=-3/2}^{3/2}\langle m\mid u\rho u^\dagger\mid u\mid\ln\langle
m\mid u\rho u^\dagger\mid m\rangle\nonumber\\
-\sum_\alpha\langle\alpha \mid u_1\rho^{3/2}(1)
u_1^\dagger\mid\alpha\rangle\ln\Big(\sum_\alpha\langle\alpha \mid
u_1\rho^{3/2}(1)
u_1^\dagger\mid\alpha\rangle\Big)\nonumber\\
-\sum_\beta\langle\alpha \mid u_2\rho^{3/2}(2)
u_2^\dagger\mid\beta\rangle\ln\Big(\sum_\beta\langle\beta \mid
u_2\rho^{3/2}(2) u_1^\dagger\mid\beta\rangle\Big). \end{eqnarray} One has the
inequality
\begin{eqnarray}\label{newmat8}
S^{(3/2)}(1)+S^{(3/2)}(2)\geq \nonumber\\
-\sum_{m=-3/2}^{3/2}\langle m\mid u_{10}\otimes u_{20}\rho
u_{10}^\dagger\otimes u_{20}^\dagger\mid m\rangle\ln\langle m\mid
u_{10}\otimes u_{20}\rho u_{10}^\dagger\otimes
u_{20}^\dagger\mid m\rangle\nonumber\\
\geq -\mbox{Tr}\,\rho\ln\rho.
\end{eqnarray}
This inequality is a new relation for the density matrix of the spin-3/2
state. In view of inequality~(\ref{newmat8}), the quantum discord is
nonnegative $D\geq 0$. The relations obtained are valid for an arbitrary
density 4$\times$4-matrix.

\subsection*{Example of qutrit state} Now we show new inequalities for qutrit
state with the density matrix
\begin{equation} \rho=
\pmatrix{\rho_{1\,1}&\rho_{1\,0}&\rho_{1~-1}\cr
\rho_{0\,1}&\rho_{0\,0}&\rho_{0~-1}\cr \rho_{-1\,1}&\rho_{-1\,0}&\rho_{-1~-1}
\cr}, \label{newmat9}
\end{equation}
considering it as a particular 4$\times$4-matrix $\rho$~\cite{JRLRVova} with
matrix elements in the fourth row and column equal to zero. In this case, the
2$\times$2-matrices associated with the 4$\times$4-matrices are
\begin{equation}\fl \rho^{(1)}(1)=
\pmatrix{\rho_{1\,1}+\rho_{0\,0}&\rho_{1\,0}\cr
\rho_{0\,1}&\rho_{-1~-1}\cr},\qquad\rho^{(1)}(2)=
\pmatrix{\rho_{1\,1}+\rho_{-1~-1}&\rho_{1\,0}\cr \rho_{0\,1}&\rho_{0\,0}\cr},
 \label{newmat10}
\end{equation}
where we used index $(1)$ to point out that the matrix is obtained for $j=1$.

We have von Neumann entropies
$S=-\mbox{Tr}\,\rho\ln\rho=-\mbox{Tr}\,\widetilde\rho\ln\widetilde\rho$,
$S_1^{(1)}=-\mbox{Tr}\,\rho_1^{(1)}\ln\rho_1^{(1)}$, and
$S_2^{(1)}=-\mbox{Tr}\,\rho_2^{(1)}\ln\rho_2^{(1)}$. The inequality $S\leq
S_1^{(1)}+S_2^{(1)}$ was obtained for the density matrix of the qutrit state
in \cite{JRLRVova}, which is a new analog of the subadditivity condition. If
we introduce tomographic probabilities corresponding to the matrix
$\widetilde\rho$ as diagonal elements of the matrix $u\widetilde\rho
u^\dagger$, i.e., $w(\alpha, u)=\langle\alpha\mid u\widetilde\rho
u^\dagger\mid\alpha\rangle$ and find the unitary 2$\times$2-matrices $u_{10}$
and $u_{20}$ diagonalizing the matrices $\rho^{(1)}(1)$ and $\rho^{(1)}(2)$,
we arrive at a stronger inequality for the qutrit density matrix, namely,
\begin{equation}
S_1^{(1)}+ S_2^{(1)}\geq H(u_{10}\otimes u_{20})\geq S, \label{newmat11}
\end{equation}
where the tomographic entropy $H(u)$ for $u=u_{10}\otimes u_{20}$ is
\begin{equation}\fl
H(u_{10}\otimes u_{20})= -\sum_\alpha\langle\alpha\mid u_{10}\otimes
u_{20}\widetilde\rho u_{10}^\dagger\otimes u_{20}^\dagger\mid\alpha\rangle \ln
\langle\alpha\mid u_{10}\otimes u_{20}\widetilde\rho u_{10}^\dagger\otimes
u_{20}^\dagger\mid\alpha\rangle.
 \label{newmat12}
\end{equation}
The quantum discord for the qutrit state reads
\begin{equation}
D^{(1)}=S_1^{(1)}+S_2^{(1)}-S-I(u_{10}\otimes u_{20}). \label{newmat13}
\end{equation}
The tomographic information $I(u_{10}\otimes u_{20})$ is determined by the
equality for an arbitrary 4$\times$4-matrix analogously to the case of
$j=3/2$. The discord for qutrit state $D^{(1)}$ is a nonnegative number. This
property is a new characteristics of quantum correlations for qutrit states.
It is clear that the procedure presented can be used to introduce quantum
discord and new inequalities for arbitrary qudit states, as well as for
multiqudit states.

\section{Conclusions}
To conclude, we point out the main results of our study.

We elaborated the method to extend all entropic and information equalities and
inequalities known for composite (both classical and quantum) systems to the
case of noncomposite and composite systems. The method is based on using an
invertible map of $N$ integers onto the pairs, triples, etc. of the integers.
A particular case of entropic inequalities like the subadditivity condition
and the strong subadditivity condition introduced for a single qudit state was
studied. We showed examples of these inequalities for qudit states with $j=2$
and 3; the subadditivity condition for qutrit state~\cite{JRLRVova} was also
presented.

We introduced the conditional probabilities and entropies for noncomposite
qudit systems and constructed chain equalities for Shannon entropy and
$q$-entropies for the qudit. Also the notion of quantum discord was given for
a single qudit. We pointed out that the physical meaning of the new
information and entropic equalities for noncomposite systems, we derived,
needs clarification though it seems that the relations, we found, correspond
to intrinsic correlations in the system, even if the system does not have the
structure of separated subsystems. These aspects of information properties of
noncomposite systems will be considered in a future publication.

\section*{Acknowledgments}
This study was initiated by the memory of our discussions with
Prof.~Rui~Vilela~Mendes and Prof.~Mary~Beth~Ruskai during the Madeira Math
Encounters XXVI (October 3--11, 2003) Quantum Information, Control and
Computing. This work was supported by the Russian Foundation for Basic
Research under Project No.~11-02-00456\_a. We are grateful to the Organizers
of the XVI Symposium ``Symmetries in Science'' (Bregenz, Austria, July 21--26,
2013) and especially to Prof. Dieter Schuch and Prof. Michael Ramek for
invitation and kind hospitality.

\end{document}